\documentclass{svproc}
\usepackage{url}

\usepackage{graphicx}
\usepackage{hyperref}
\usepackage{color}

\usepackage{hyperref}
\hypersetup{
    colorlinks=true,
    citecolor=blue,
    filecolor=black,
    linkcolor=black,
    urlcolor=gray
}
\usepackage{footmisc}

\usepackage{amsmath}
\usepackage[svgnames,table]{xcolor}
\usepackage{enumitem}
\usepackage{fontawesome}
\usepackage{multirow}
\usepackage{listings}
\usepackage{mdframed}
\usepackage{framed}
\usepackage{soul}
\usepackage{multicol}
\usepackage{footmisc}
\usepackage[margin=1in]{geometry}
\usepackage{subcaption}

\usepackage{tabularx}
\usepackage{booktabs}
\usepackage{array}
\newcolumntype{Y}{>{\raggedright\arraybackslash\hspace{0pt}}X}

\usepackage{siunitx}        
\newcolumntype{L}{>{\raggedright\arraybackslash}X}

\lstdefinelanguage{Cypher}{
    morekeywords={
        MATCH, RETURN, ORDER, BY, LIMIT, AS, COUNT, DESC
    },
    sensitive=true,
    morecomment=[l]{//},
    morestring=[b]",
}

\newcommand{\nd}{\vspace{1mm}\noindent}

\newenvironment{custombox}{\smallskip\begin{mdframed}[linewidth=1pt,innerleftmargin=4pt, innerrightmargin=4pt, innertopmargin=3pt, innerbottommargin=3pt, nobreak=true]}{\end{mdframed}\smallskip}

\begin{document}
\mainmatter              
\title{Structural and Connectivity Patterns in the Maven Central Software Dependency Network}
\titlerunning{Structural and Connectivity Patterns} 
%
\author{Daniel Ogenrwot \and John Businge \and Shaikh Arifuzzaman}
\authorrunning{D. Ogenrwot et al.} 
%
\tocauthor{Daniel Ogenrwot,John Businge and Shaikh Arifuzzaman}
\institute{University of Nevada Las Vegas, Las Vegas NV 89154, USA,\\
\email{ogenrwot@unlv.nevada.edu, \{john.businge, shaikh.arifuzzaman\}@unlv.edu}
}

\maketitle              

\begin{abstract}
Understanding the structural characteristics and connectivity patterns of large-scale software ecosystems is critical for enhancing software reuse, improving ecosystem resilience, and mitigating security risks. In this paper, we investigate the \texttt{Maven Central} ecosystem, one of the largest repositories of Java libraries, by applying network science techniques to its dependency graph. Leveraging the Goblin framework, we extracted a sample consisting of the top 5,000 highly connected artifacts based on their degree centrality and then performed breadth-first search (BFS) expansion from each selected artifact as a seed node, traversing the graph outward to capture all libraries and releases reachable those seed nodes. This sampling strategy captured the immediate structural context surrounding these libraries resulted in a curated graph comprising of 1.3 million nodes and 20.9 million edges. We conducted a comprehensive analysis of this graph, computing degree distributions, betweenness centrality, PageRank centrality, and connected components graph-theoretic metrics. Our results reveal that \texttt{Maven Central} exhibits a highly interconnected, scale-free, and small-world topology, characterized by a small number of infrastructural hubs that support the majority of projects. Further analysis using PageRank and betweenness centrality shows that these hubs predominantly consist of core ecosystem infrastructure, including testing frameworks and general-purpose utility libraries. While these hubs facilitate efficient software reuse and integration, they also pose systemic risks; failures or vulnerabilities affecting these critical nodes can have widespread and cascading impacts throughout the ecosystem. 
\keywords{Software dependencies, Dependencies network analysis, Maven central repository, Software security, Software ecosystems, Network data mining}
\end{abstract}
\section{Introduction} \label{sec:intro}

The scale and complexity of modern software ecosystems have expanded rapidly due to the adoption of modular development practices and widespread reuse of open-source libraries~\cite{Mens2014,Decan2016,MANIKAS201684,Bogart2016}. \texttt{Maven Central}, one of the most influential repositories for Java artifacts, plays a pivotal role in contemporary software engineering by serving as a centralized hub through which millions of developers access and manage dependencies~\cite{RAEMAEKERS2017140}. While this ecosystem prompt reuse, accelerates development, and promotes innovation, it also creates significant complexity in understanding the structural dynamics, connectivity patterns, and systemic risks inherent within the ecosystem \cite{Abate2009}. For example, Bavota et al.~\cite{Bavota2015} analyzed the impact of library changes on client systems and highlighted the widespread ripple effects caused by even minor updates, emphasizing the challenge of managing implicit and transitive dependencies. Understanding these properties is significant for modern software development. The structural characteristics of software ecosystems have direct implications for software engineering practices, ecosystem sustainability, and software security \cite{Decan2016,Decan2019,Mens2014,johannes2022}. For example, the existence of highly central artifacts suggests focal points for optimization, but also highlights critical points of failure. It also introduces considerable systemic complexity, making it increasingly difficult to reason about transitive dependencies, hidden couplings, and the potential for cascading failures~\cite{Abate2009,Decan2019}. These studies emphasize the need to move beyond localized reasoning to account for the broader structural properties of software ecosystems. 

The importance of understanding such structural complexity is illustrated by real-world incidents such as the \textit{Log4Shell} vulnerability in the \texttt{log4j} library~\cite{hiesgen:2024:log4j}. This widely used logging utility, embedded transitively in countless Java projects, exposed a critical zero-day vulnerability (CVE-2021-44228)~\footnote{\url{https://nvd.nist.gov/vuln/detail/cve-2021-44228}} that affected millions of systems globally. Despite many projects not directly depending on \texttt{log4j}, their inclusion of libraries that in turn depended on it made them vulnerable. This event highlighted the urgent need to comprehend not only direct dependencies, but also the deeper structural and connectivity patterns of the ecosystem to evaluate systemic risk, and develop robust mitigation strategies.

The field of software ecosystem analysis has seen significant growth, especially as open-source repositories like \texttt{Maven Central}~\footnote{\url{https://mvnrepository.com/repos/central}}, \texttt{npm}~\footnote{\url{https://www.npmjs.com/}}, and \texttt{PyPI}~\footnote{\url{https://pypi.org/}} have become essential backbones of modern software development. Prior research has explored various aspects of software ecosystems, including semantic versioning~\cite{Decan2021,Wenke2023} and the impact of breaking changes~\cite{RAEMAEKERS2017140}, strong dependency relationships~\cite{Abate2009}, the effects of dependency smells on software quality~\cite{JOLAK2025112382}, the macro-level evolution of large software compilations~\cite{Gonzalez-Barahona2009}, vulnerability tracking~\cite{yang2025tracing,Kumar2024,Zeyang2024,johannes2022,jukka2018}, dependency management tools, and localized impact assessments of specific libraries~\cite{johannes2022} or versions~\cite{yoshioka2025developers}. Tools such as Dependabot and Renovate~\cite{runzhi2023} help developers manage and update their local dependencies efficiently~\cite{Chen2015}. However, most existing studies concentrate on localized or pairwise relationships, such as whether a project is up to date or if a particular dependency contains known vulnerabilities~\cite{Kula2018,Zeyang2024}. These micro-level analyses, while essential, often fall short in addressing the ecosystem's global structure, resilience, and the role of infrastructural libraries in propagating risk~\cite{Decan2019}. For example, Zerouali et al.~\cite{Zerouali:2018} explored the adoption of library updates and found that developers frequently avoid upgrades due to fear of breaking changes, leading to the accumulation of technical lag. While informative, their study primarily focuses on the relationship between a project and its direct dependencies. Similarly, Abdalkareem et al.~\cite{Abdalkareem:2017} studied the impact of utility libraries on software quality, showing that certain libraries disproportionately influence maintainability and defect proneness. However, they do not consider how these central libraries are embedded within the broader network or how their failure might affect global connectivity. These findings highlight the necessity for a macro-structural perspective that captures the emergent properties of the ecosystem and enables reasoning about cascading effects, centrality, and resilience.

To address this gap, we adopt a macro-structural perspective on \texttt{Maven Central} by modeling its dependency relationships as a directed network. Using techniques from network science, we construct and analyze a comprehensive dependency graph that captures over 1.3 million artifacts and 20.9 million directed edges. We employ graph-theoretic metrics \cite{arifuzzaman2019scalable,sattar2025dyg,sattar2020data,sattar2023exploring,sattar:2019,tkdd-arif20} such as degree distributions, PageRank, betweenness centrality, and connected component analysis to uncover hidden patterns, identify structurally critical nodes, and assess the ecosystem's connectivity and resilience to targeted failures. This global lens allows us to answer questions such as: How interconnected is \texttt{Maven Central}? Which libraries serve as infrastructural backbones? How vulnerable is the ecosystem to cascading failures triggered by the removal of central artifacts? To the best of our knowledge, this is the largest scale-free and small-world topological analysis conducted on \text{Maven Central} to date, combining both global (PageRank) and path-sensitive (betweenness) metrics across millions of nodes. This paper makes the following contributions:
\begin{itemize}
    \item We construct a large-scale dependency graph of the \texttt{Maven Central} ecosystem, comprising over 1.3 million nodes and 20.9 million directed edges, capturing transitive and direct dependency relationships.
    \item We apply graph-theoretic and network science techniques to quantify structural properties using degree distributions, PageRank, betweenness centrality, and connected components metrics.
    \item We identify critical infrastructural libraries whose central roles in the network pose both opportunities for optimization and potential sources of systemic fragility.
    \item We release a fully reproducible replication package, including the dataset, preprocessing scripts, and analysis notebooks, to facilitate future research and validation~\cite{replication-package:2025}.
\end{itemize}

\section{Related Work}\label{sec:relatedwork}
This section reviews foundational and recent research relevant to our investigation of \texttt{Maven Central}. First, we examine the structure and evolution of software ecosystems and their dependency networks. Next, we explore the application of graph mining and network analysis in software engineering domain. We then discuss recent studies specifically focusing on the \texttt{Maven Central} dependency graph. Finally, we identify persisting research gaps in the field. 

\textbf{Software Ecosystems and Dependency Networks.}  
Software ecosystems comprise interrelated software projects that evolve together within a shared technical and organizational context, shaped by dependencies, governance structures, and social interactions among stakeholders~\cite{Mens2014,MANIKAS201684}. Foundational models of these ecosystems highlight the co-evolutionary interplay between technical architectures and community practices as critical forces driving their sustainability and growth~\cite{Jansen2009,bosch2009software}.  
Building upon these theoretical insights, Businge et al.~\cite{busing:2022:reuse} provided empirical evidence by investigating how such dynamics unfold in practice, particularly through the lens of reuse and maintenance strategies among divergent forks across multiple ecosystems. Their broader investigations have also addressed API usage trends~\cite{businge:2013:api,businge:2015:api} and software variants~\cite{businge2023:analyze,variant:2022:saner,ogenrwot:2025:repatch,pareco:2022}, all of which contribute to the understanding of how ecosystems accommodate change, specialization, and reuse. \texttt{Maven Central}, a prominent repository for Java artifacts, offers a compelling case study of modern software supply chains due to its extensive reuse patterns, complex versioning schemes, and decentralized governance. Decan et al.~\cite{Decan2019} analyzed the dependency network in seven software ecosystems. Their work highlighted that not only the number of dependencies but also their structural arrangement determine the resilience of the ecosystem to breaking changes. More recently, studies have investigated the socio-economic dimensions of dependency networks. Saied et al.~\cite{SAIED2018164} presented a usage-pattern mining approach on \texttt{Maven Central} that automatically identifies cohesive clusters of libraries used together by thousands of client projects, highlighting common co-dependency structures and opportunities for improved recommendation and tooling support. Similarly, Kikas et al.~\cite{kikas:2017} showed that isolated clusters often correspond to experimental or short-lived modules, which may be pruned or deprecated over time to reduce technical debt. Understanding the topology and dynamics of dependency graphs provides actionable insights for repository governance, dependency management tooling, and ecosystem health monitoring.

\textbf{Graph Mining and Network Analysis in Software Ecosystems.}  
Graph mining plays a fundamental role in extracting insights from complex network structures~\cite{abdelhamid2014cinet,arifuzzaman2012patric,arifuzzaman:2024,faysal2021hypc,shafin2025faster}. The application of network science and graph mining techniques to model and analyze software ecosystems is well established. Valverde and Solé~\cite{valverde2003hierarchical} were among the first to characterize software dependency structures as scale-free networks, revealing non-random hierarchical topologies. Building on this foundation, Jansen and Brinkkemper~\cite{Jansen2009} and Cosentino et al.~\cite{cosentino2015} employed community detection, centrality, and clustering techniques to study software architecture, module cohesion, and ecosystem evolution over time. In the context of \texttt{Maven Central}, structural metrics such as modularity, path length, and node influence have proven critical for understanding system-level behavior and propagation dynamics. Recent research has leveraged temporal graph analysis and resilience modeling to uncover fragility points and assess robustness under evolving dependency loads~\cite{Decan2019}. Several studies focus on the evolution, risk, and hidden complexity of dependency networks. Kula et al.~\cite{Kula2018} investigated library migration practices and identified developer hesitancy in updating dependencies, often driven by concerns over breaking changes and integration effort. Bavota et al.~\cite{Bavota2015} explored how API changes ripple through ecosystems, exposing the costs and fragility associated with software reuse. In a complementary study, Decan et al.~\cite{Decan2016} differentiated between direct and transitive dependencies, revealing that the majority of a project’s dependency footprint lies within transitive layers, often outside the developer’s immediate awareness or control. These studies highlight the growing complexity of modern software ecosystems and emphasize the technical debt and sustainability risks posed by unmanaged or opaque dependency structures.

\textbf{Recent Studies on the \texttt{Maven Central} Repository.} Building on this, recent large-scale analyses have begun to explore previously overlooked components of ecosystems. Notably, Shanto et al. \cite{shanto2025dependency} conducted a comprehensive study of 658,078 Maven artifacts, revealing that 15.4\% of artifacts had no incoming dependencies (in-degree = 0), termed `independent artifacts'. These findings position independent artifacts as safer, self-contained alternatives, though the study also flagged maintainability challenges. Furthermore, Yang-Smith and Abdellatif~\cite{yang2025tracing} analyzed 3,362 CVEs in Maven to explore the dynamics of vulnerability disclosure and mitigation across parent and dependent packages. A notable finding was the prevalence of ``Publish-Before-Patch'' scenarios, where severe vulnerabilities received faster patching post-disclosure, reducing response times from 151 to 78 days. Chowdhury et al.~\cite{chowdhury2025insights} investigated dependency maintenance trends in the Maven ecosystem, focusing on issues such as outdated dependencies, missed releases, and the challenges posed by complex dependency structures. Their quantitative analysis revealed that projects with fewer dependencies tend to have a higher incidence of missed releases, potentially undermining software quality and stability. Conversely, dependencies in the latest releases often exhibited positive freshness scores, suggesting improved dependency management practices in newer versions. Shafin et al.~\cite{shafin2025faster} explored the interplay between release practices, dependency freshness, and security in Maven artifacts. 
Their findings revealed that artifacts releasing updates more rapidly and consistently tended to remain up to date for longer periods and exhibited fewer vulnerabilities in their dependency chains.

\textbf{Summary and Research Gaps.}
Besides the aforementioned studies, several studies have investigated the \texttt{Maven Central} repositories to understand how developer interact with deprecated library versions \cite{yoshioka2025developers}, vulnerability management \cite{rabbi2025chasing,nachuma2025decoding,rabbi2025understanding,Kumar2024,johannes2022}, among many others. However, there is still limited understanding of the holistic structural and connectivity properties of large-scale ecosystems like \texttt{Maven Central}. While prior work~\cite{Kula2018,Decan2016,Bavota2015} has focused on localized or version-level analysis of dependency networks; our study provides a holistic, macro-structural view, quantifying centrality and connectivity across millions of artifacts and highlighting systemic fragility in ways not previously explored at this scale. Our work contributes to filling this gap by performing a comprehensive graph-based analysis of \texttt{Maven Central}’s dependency network, characterizing its modularity, centralization, and connectivity evolution. In doing so, we aim to enrich the theoretical and practical understanding of software ecosystem resilience and architecture.

\section{Methodology}\label{sec:methodology}
This section presents comprehensive research questions and the methodology pipeline adopted in this study.

\subsection{Research Questions}
\begin{itemize}[leftmargin=*]
    \item \textbf{RQ1:} \textit{How do metrics such as degree distribution characterize the dependency graph? Is the graph scale-free, small-world, or does it exhibit other known graph structures?}
    This research question can be divided into two parts. The first part examines how connections are distributed across the graph, identifying whether the network is dominated by a few highly connected hubs or is more evenly distributed. The second part investigates the overall architecture of the ecosystem to determine if it follows patterns seen in other complex systems. This is important because scale-free or small-world networks are known to be robust against random failures but vulnerable to targeted attacks, and understanding this helps assess the ecosystem's resilience and guides risk management strategies. 
    \item \textbf{RQ2:} \textit{Are certain types of projects more likely to be central (hubs) or peripheral (leaves) in the graph structure?} This question explores whether specific categories of projects (e.g., frameworks, libraries, utilities) systematically occupy key network positions. This is important because identifying which project types act as critical hubs can inform decisions on maintenance priorities, security audits, and resource allocation.
    \item \textbf{RQ3:} \textit{Is the graph made up of connected components with no relationship between them?} This question asks whether the ecosystem is fragmented into isolated subgraphs or forms a cohesive whole. This is important because a cohesive network suggests a tightly integrated ecosystem with widespread interdependence, while fragmentation could signal isolated communities or unused modules, shaping strategies for ecosystem improvement and outreach.
\end{itemize}

\subsection{Methodology Pipeline}\label{sec:method}
\begin{figure}[htbp]
    \centering
    \includegraphics[width=\linewidth]{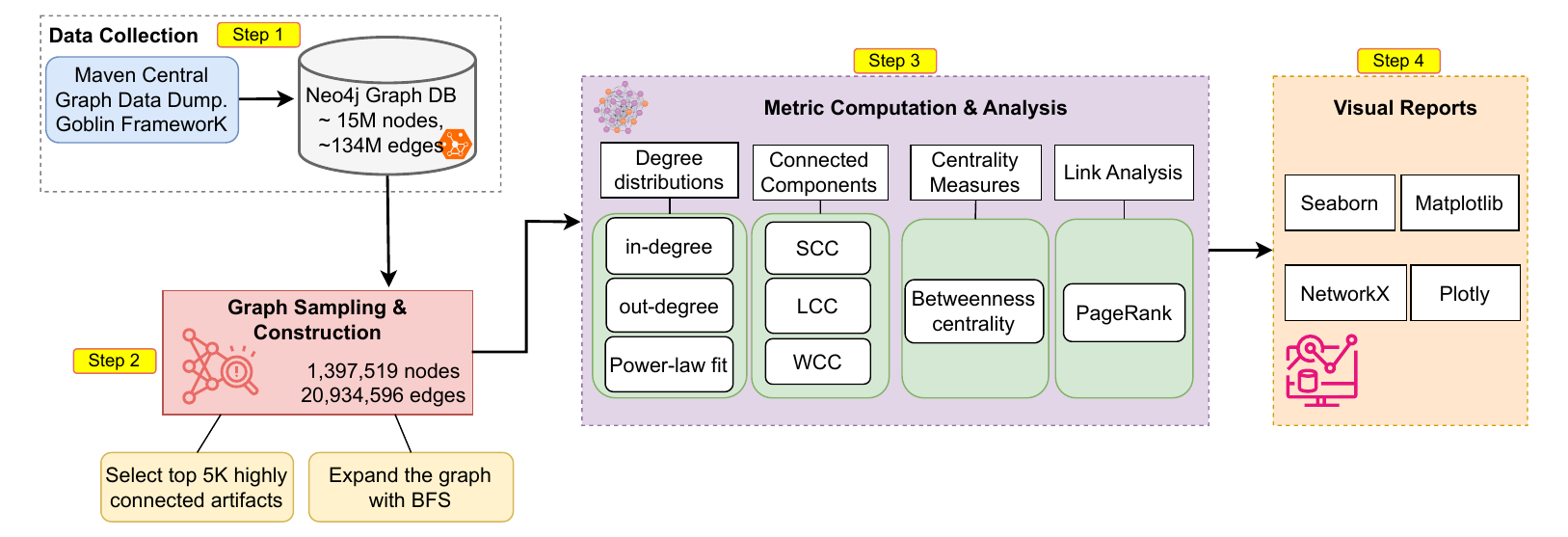}
    \caption{Overview of the methodology pipeline.}
    \label{fig:method}
\end{figure}

An overview of the methodology pipeline is depicted in Figure~\ref{fig:method}. The process is organized into four main steps: \textit{Step 1} describes the data collection process, detailing how we extracted and prepared the dependency information from \texttt{Maven Central}. \textit{Step 2} explains the graph sampling and reconstruction techniques used to build a representative and analyzable network. \textit{Step 3} focuses on the computation of structural metrics using graph-theoretic algorithms. Finally, \textit{Step 4} involves visual and exploratory analysis to interpret the network’s topology and identify key structural patterns.

\nd \textbf{Step 1: Data collection}. The study uses the \texttt{Maven Central} dependency graph provided through the Goblin framework and a Neo4j-based graph database released as part of the 2025 Mining Software Repository (MSR) mining challenge. Specifically, we used the dataset version \texttt{with\_metrics\_goblin\_maven\_30\_ 08\_24.dump}, dated August 30, 2024 \cite{GoblinDataset2024}. The dataset includes approximately 15 million nodes, comprising over 600,000 distinct libraries and more than 14 million releases. The graph also contains approximately 134 million edges, including over 119 million dependency edges and over 14 million versioning edges. The graph consists of two primary node types: \textit{Artifact nodes}, which represent software components such as libraries, frameworks, or tools, and \textit{Release} nodes, which denote specific published versions of these artifacts. Each release node includes metadata such as the version identifier and its publication timestamp (in Unix format). Artifacts are linked to their respective releases via a one-to-many relationship, reflecting multiple releases per artifact. Additionally, the graph contains \textit{dependency edges}, which establish many-to-many relationships from release nodes to other artifact nodes, capturing both direct and transitive dependencies among components. This rich and large-scale dataset enables a comprehensive analysis of the structural and connectivity properties of the \texttt{Maven Central} ecosystem.

\nd \textbf{Step 2: Graph Sampling and Construction}.  
To extract a representative and computationally manageable subgraph from the full \texttt{Maven Central} dependency graph, we selected the top 5,000 most highly connected artifacts based on their incoming degree. This strategy is motivated by the scale-free nature of the software dependency ecosystems, where a small subset of central nodes contributes disproportionately to overall connectivity and dependency propagation~\cite{Decan2016,barabasi1999emergence}. We used Cypher query executed against the Neo4j database (Listing~\ref{lst:top5k}). This query can be easily adapted to retrieve any number $N$ of top-ranked artifacts, depending on the desired sampling scope.

\begin{lstlisting}[label={lst:top5k},caption={Cypher query to identify the top 5,000 highly connected artifacts in the Maven Central graph.}]
MATCH (r:Release)-[:dependency]->(a:Artifact)
RETURN a.id AS artifactId, COUNT(*) AS numIncoming
ORDER BY numIncoming DESC
LIMIT 5000;
\end{lstlisting}

\noindent To capture the immediate structural context surrounding these artifacts, we applied a breadth-first search (BFS) expansion from each selected artifact as a seed node, traversing the graph outward to a depth of two. This sampling strategy ensures that all releases associated with each artifact, as well as all artifacts that depend on any of these releases are included. To avoid redundancy, we implemented a node visitation check, ensuring that each artifact or release node is only included once in the sampled graph. We then constructed a directed graph using NetworkX \cite{platt2019network}, modeling two types of relationships: (i) artifact-to-release edges representing the release lineage of an artifact, and (ii) release-to-dependent-artifact edges capturing the dependency relations. The resulting graph comprises 1,397,519 nodes (libraries and releases) and 20,934,596 edges, providing a rich and a structurally diverse dataset suitable for subsequent metric computation and visualization.

\nd \textbf{Step 3: Metrics Computation.}
To characterize the structural properties of the sampled \texttt{Maven Central} dependency network, we modeled the ecosystem (subgraph generated constructed in \textit{Step 2}) as a directed graph $ G = (V, E) $, where nodes $ V $ represent software artifacts (libraries and releases) and edges $ E \subseteq V \times V $ capture directed dependency relations. We computed a suite of graph-theoretic metrics to analyze this network, including degree distributions \cite{tkdd-arif20}, centrality measures \cite{faysal2023fast}, and connected components \cite{faysal2021hypc,sattar2025dyg}. For each node $ v \in V $, we calculated the in-degree $ d^{-}(v) $ and out-degree $ d^{+}(v) $, defined respectively in equation~\ref{eqn:degree}, Where $ d(v) = d^{-}(v) + d^{+}(v) $ denotes the total degree.

\begin{equation}\label{eqn:degree}
d^{-}(v) = |\{u \in V : (u, v) \in E\}|, \quad d^{+}(v) = |\{u \in V : (v, u) \in E\}|,
\end{equation} 

\nd These metrics provide foundational insights into module popularity and dependency sprawl. We also computed betweenness centrality to quantify the control a node exerts over information flow, using the standard formulation~\ref{eqn:betweenness} where $ \sigma_{st} $ is the number of shortest paths between nodes $ s $ and $ t $, and $ \sigma_{st}(v) $ is the number of path that pass through  node $ v $~\cite{liu:2025:novel}.

\begin{equation}\label{eqn:betweenness}
C_B(v) = \sum_{\substack{s, t \in V \\ s \neq v \neq t}} \frac{\sigma_{st}(v)}{\sigma_{st}},
\end{equation}

\nd To measure influence via recursive link analysis, we employed PageRank centrality, computed as shown in equation~\ref{eqn:pagerank},where $ \alpha \in (0,1) $ is a damping factor (typically set to 0.85) and $ \text{Pre}(v) $ denotes the set of predecessors of node $ v $.

\begin{equation}\label{eqn:pagerank}
PR(v) = \frac{1 - \alpha}{|V|} + \alpha \sum_{u \in \text{Pre}(v)} \frac{PR(u)}{d^{+}(u)},
\end{equation}

\noindent In addition, we analyzed largest connected components (LCC), weakly connected components (WCC) and strongly connected components (SCC) to assess ecosystem modularity and fragmentation; the former considers connectivity when edge direction is ignored, while the latter requires mutual reachability under directed paths. All metrics were computed using the \texttt{NetworkX}~\cite{platt2019network} Python library, chosen for its robustness in handling large directed graphs. This combination of metrics was selected based on both theoretical foundations and empirical precedent to provide a comprehensive understanding of topological prominence, dependency cohesion, and network resilience. These insights directly support our broader investigation into centralization trends, single points of failure, and the structural fragility of software ecosystems. 

\nd \textbf{Step 4: Visual Reports}.
To facilitate the interpretation and communication of our findings, we generated a series of visualizations using NetworkX, Matplotlib, Seaborn and Plotly python packages. These tools enabled us to produce network diagrams, degree distribution plots, and clustering visualizations. The resulting figures provide both quantitative summaries and qualitative insights into the structural characteristics of the ecosystem, ensuring that complex patterns are presented in an interpretable and accessible manner.


\section{Results \& Discussion}\label{sec:results}
\subsection*{RQ1: How do metrics such as degree distribution characterize the dependency graph? Is the graph scale-free, small-world, or does it exhibit other known graph structures?}

\begin{figure}[htbp]
    \centering
    \begin{subfigure}[b]{0.49\linewidth}
        \includegraphics[width=\linewidth]{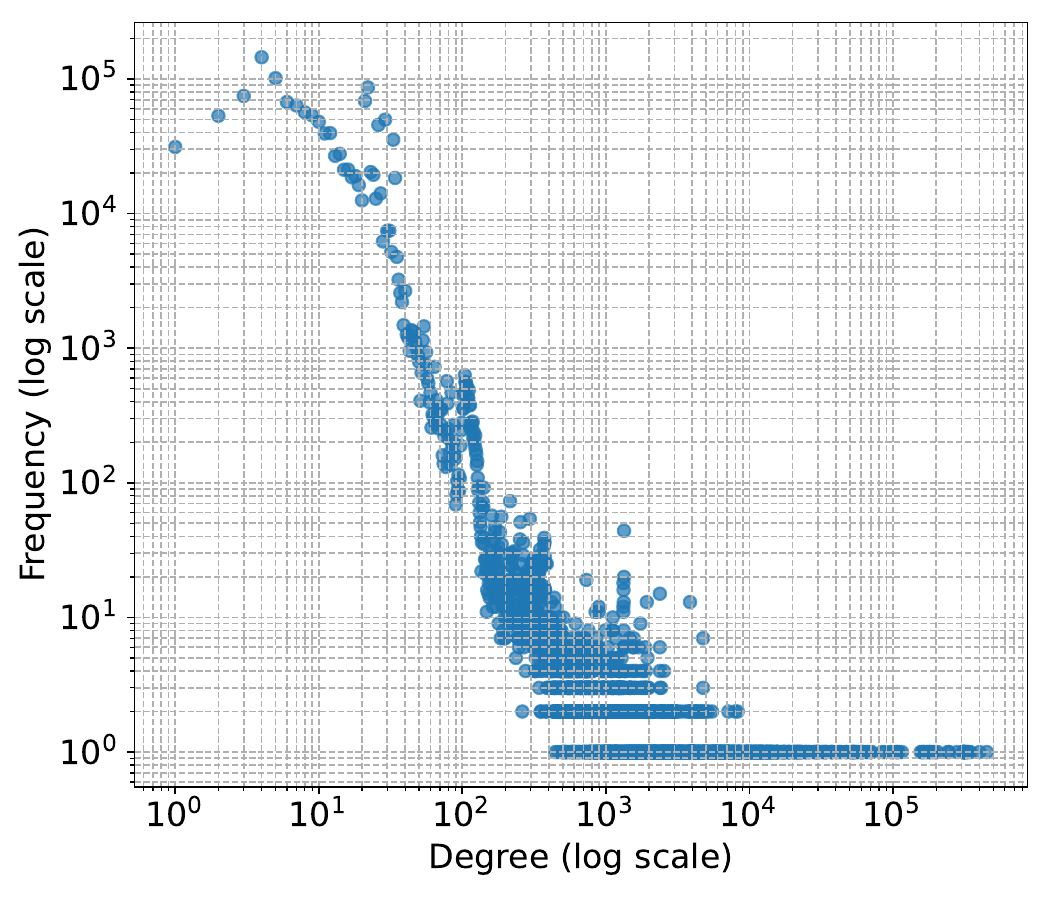}
        \caption{Out- and in-degree distributions of all artifacts.}
        \label{fig:distribution-all}
    \end{subfigure}
    \hfill
    \begin{subfigure}[b]{0.49\linewidth}
        \includegraphics[width=\linewidth]{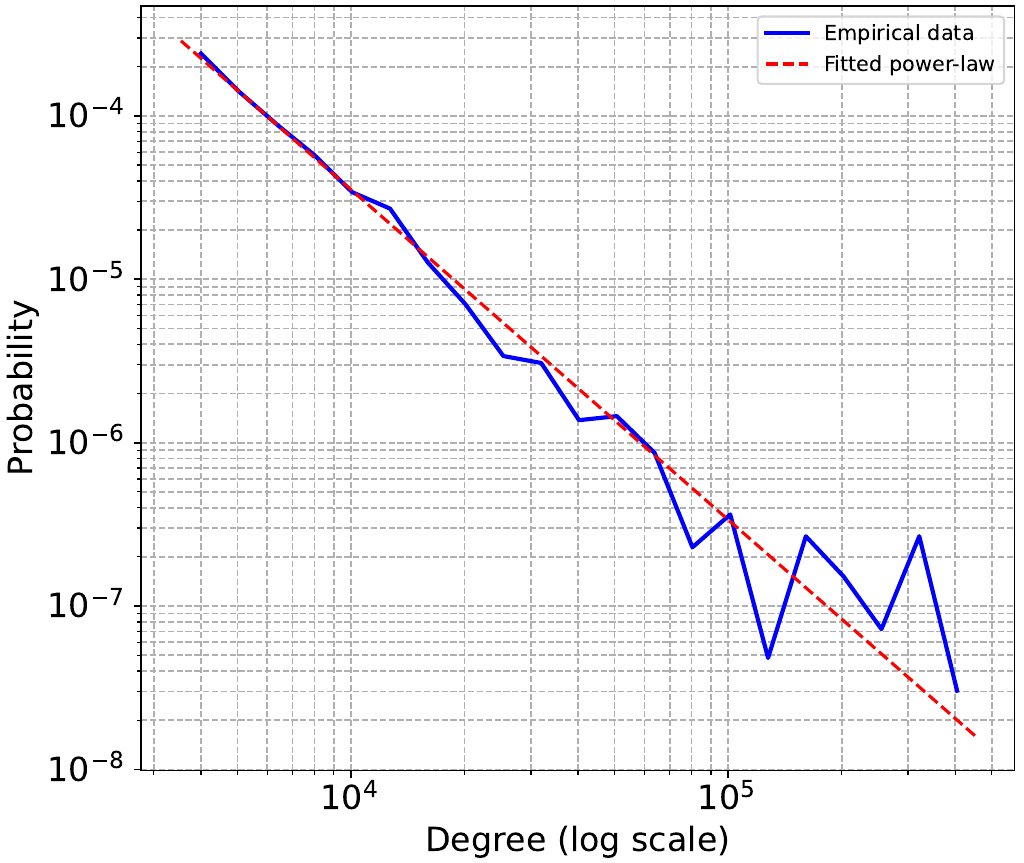}
        \caption{Sampled graph degree distribution with power-law fit.}
        \label{fig:powerlaw-fit}
    \end{subfigure}
    \caption{Comparison of degree distributions in the Maven dependency graphs. (a) shows full-graph out- and in-degree distributions. (b) highlights the power-law fit for the sampled graph.}
    \label{fig:degree-distribution-powerlawfit}
\end{figure}

To understand the structural properties of the \texttt{Maven Central} dependency graph, we first computed the overall degree distribution using the sampled dataset described in Section~\ref{sec:methodology}. As shown in Figure~\ref{fig:distribution-all}, the distribution is highly skewed and exhibits a heavy-tailed pattern, indicating a small number of nodes with extremely high connectivity and a majority with minimal connections. This reflects a centralized reuse structure commonly observed in large-scale software ecosystems. To investigate whether this distribution follows a power-law (a hallmark of scale free networks), we used the \texttt{powerlaw} Python library to fit a power-law model to the sampled data. The resulting fit (Figure~\ref{fig:powerlaw-fit}) confirms power-law behavior in the tail of the distribution, consistent with scale-free topologies. Visual inspection of the log-log plot further supports this, as the empirical curve aligns closely with the theoretical power-law line.

\begin{figure}[htbp]
    \centering
    \begin{subfigure}[b]{0.49\linewidth}
        \includegraphics[width=\linewidth]{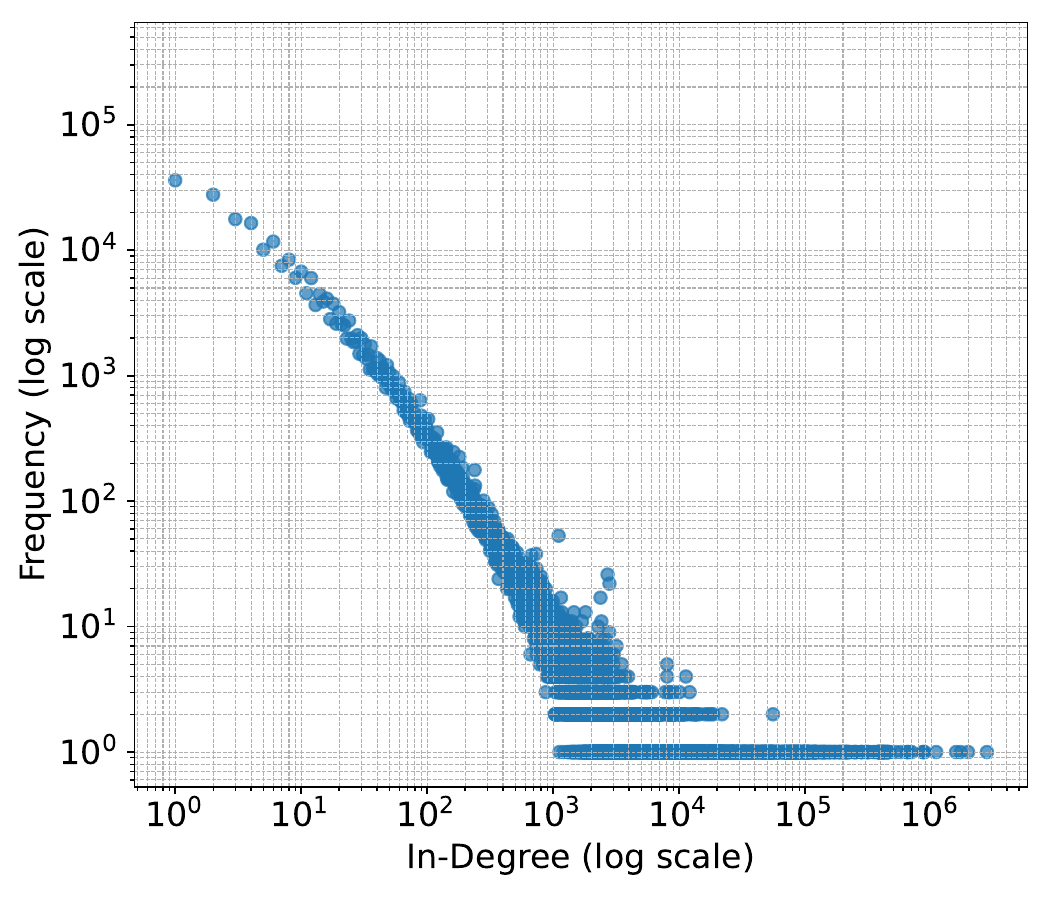}
        \caption{Libraries (in-degree)}
        \label{fig:distribution-libraries}
    \end{subfigure}
    \hfill
    \begin{subfigure}[b]{0.49\linewidth}
        \includegraphics[width=\linewidth]{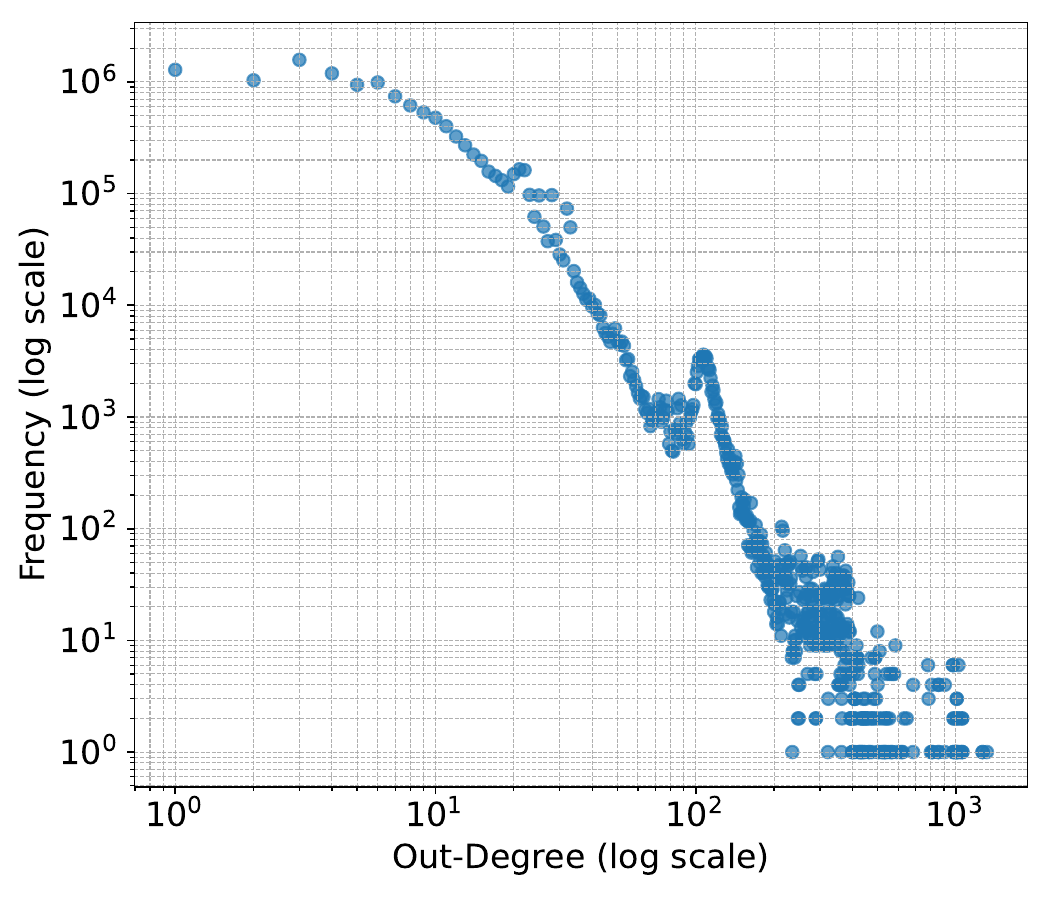}
        \caption{Releases (out-degree)}
        \label{fig:distribution-releases}
    \end{subfigure}
    \caption{Degree distributions in the Maven dependency graphs, plotted in log-log scale. (a) In-degree distributions of all libraries in the Maven dependency graph., and (b) Out-degree distributions of all releases in the Maven dependency graph.}
    \label{fig:degree-distributions-comparison}
\end{figure}

To further understand how this global structure manifests at the component level, we next examined the in-degree and out-degree distributions of libraries and releases, respectively. As shown in Figure~\ref{fig:degree-distributions-comparison}, the in-degree distribution of libraries and the out-degree distribution of releases both follow heavy-tailed patterns. This supports the finding that a small number of core libraries are heavily reused across many releases, whereas most releases depend on relatively few other components.

\noindent Overall, these results demonstrate that the \texttt{Maven Central} ecosystem exhibits both scale-free and small-world properties. The scale-free structure, driven by a few highly connected infrastructural nodes, enables efficient reuse but also introduces systemic risks such as cascading failures or vulnerability propagation if these hubs are compromised.

\subsubsection{Discussion and Implications.}
Our analysis reveals that the \texttt{Maven Central} dependency graph follows a power-law degree distribution and exhibits small-world characteristics. These structural properties have clear implications for how the ecosystem functions and where it can fail. From a network science perspective, the presence of a power-law distribution suggests that \texttt{Maven Central} behaves like a scale-free network. In such networks, a small number of nodes, called \textit{hubs}, have a very large number of connections, while most nodes have relatively few. This mirrors the structure we observe, where a few widely reused libraries support a large number of dependent projects. These hubs often emerge due to preferential attachment, where popular libraries are more likely to attract additional dependents over time. In practice, this structure benefits the ecosystem since developers can rely on well-tested, widely adopted libraries to accelerate development and reduce duplication. The small-world property of the network means that any two artifacts are typically connected through just a few intermediate dependencies. This property enhances dependency resolution and encourages modular yet interconnected designs. 

However, these same features also create risks because large proportion of the ecosystem depends on a few central libraries. Any issues with those hub such as bugs, security vulnerabilities, or a lack of maintenance can have widespread effects. Failures in these components may break builds, introduce security holes, or cause large-scale disruptions. Incidents like the left-pad removal in \texttt{npm}~\cite{Abdalkareem:2017} or the \textit{Log4Shell} vulnerability in \texttt{log4j}~\cite{hiesgen:2024:log4j} show how fragile these systems can become when critical libraries are affected. To address these risks, dependency management tools should go beyond checking for outdated or vulnerable versions. They should also consider the network role of each dependency. For example, tools could warn developers when they are about to depend on a high-centrality artifact, or suggest alternative libraries to reduce concentration. Ecosystem maintainers can use this structural information to identify libraries that need stronger governance, additional maintainers, or dedicated funding

\begin{custombox}
\faLightbulbO \hspace{0.1mm} \emph{\textbf{Summary - RQ1}:
The \texttt{Maven Central} dependency graph exhibits a highly skewed, heavy-tailed degree distribution, with a small number of hubs possessing very high connectivity and a large number of nodes with minimal connections. The degree distribution follows a power-law behavior, confirming that the ecosystem exhibits scale-free network properties. The graph also demonstrates small-world characteristics. These structural patterns reflect a centralized reuse model, where key artifacts serve as infrastructural hubs supporting many dependent projects.
}
\end{custombox}

\subsection*{RQ2: Are certain types of projects more likely to be central (hubs) or peripheral (leaves) in the graph structure?}

\begin{figure}[!h]
  \centering
  \includegraphics[width=0.9\linewidth]{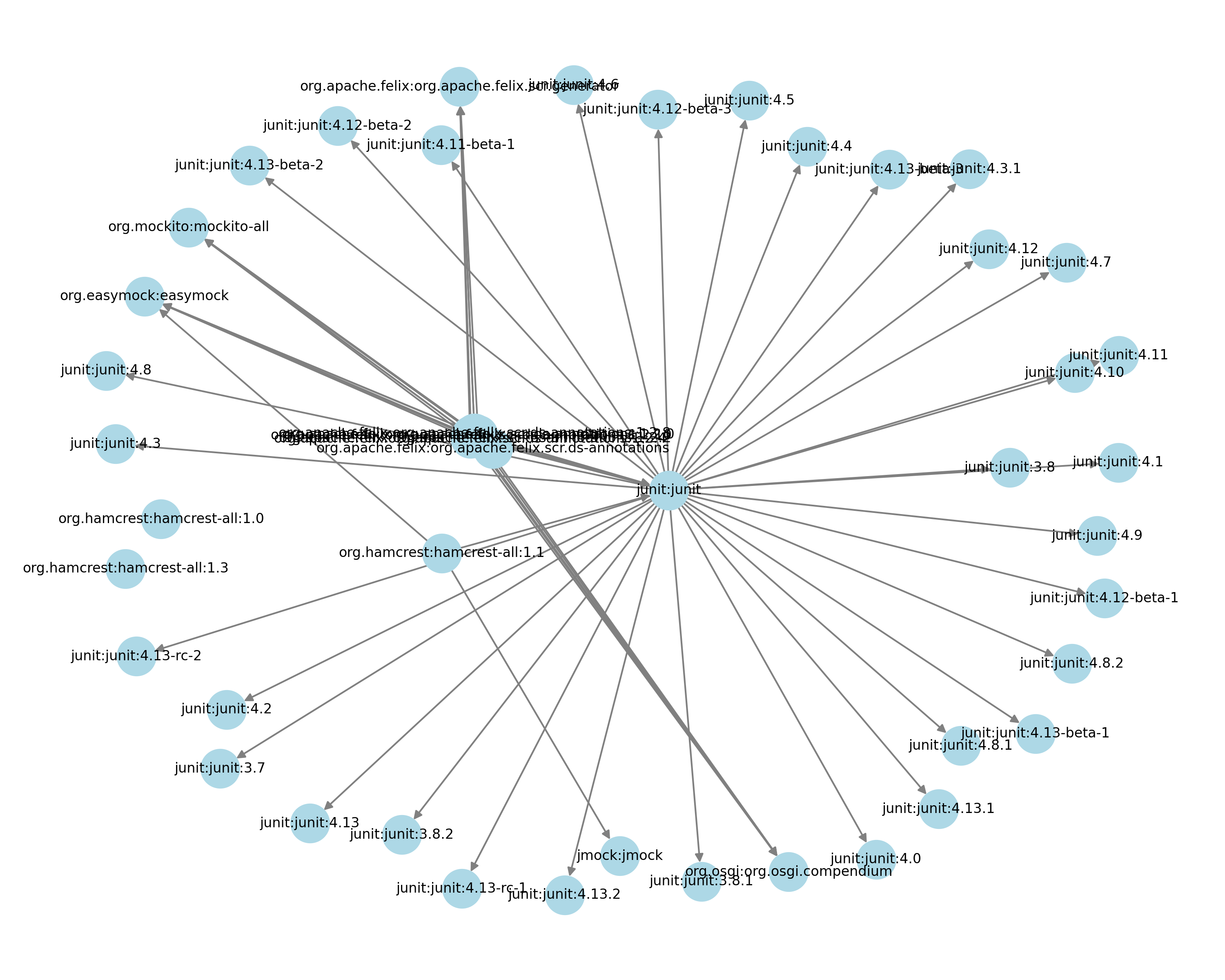}
  \caption{Network visualization of the top 10 artifacts by PageRank in the Maven Central dependency graph. Each hub is connected to its immediate neighbors, including all release versions and dependent artifacts. 
  }
  \label{fig:pagerank-top10}
\end{figure}

To identify the most influential nodes in the Maven Central dependency graph, we computed the PageRank centrality for all nodes in the sampled graph. Table~\ref{tab:pagerank} presents the top 10 nodes ranked by their PageRank score, representing the primary hubs of the ecosystem. Figure~\ref{fig:pagerank-top10} shows the network visualization of the ten artifacts with the highest PageRank scores and their immediate neighbors. Our analysis reveals that the most central nodes predominantly consist of core infrastructure libraries. Notably, the \texttt{org.apache.felix:org.apache.felix.scr.ds\-annotations} artifact and its versions occupy the highest ranks, indicating their critical role in defect detection and metadata analysis across the ecosystem. These nodes consistently show the highest PageRank scores, reflecting their significant influence within the dependency network. In addition, widely used testing frameworks such as \texttt{junit:junit} and the \texttt{org.hamcrest: hamcrest-all} family also appear among the top hubs. These libraries are essential components for quality assurance and testing activities, reaffirming their importance not only within project-level practices but also as ecosystem-wide enablers of software reliability and verification. These findings align with existing knowledge that infrastructural and quality-assurance libraries tend to emerge as key hubs in software ecosystems \cite{Decan2016,Mens2014,Kula2018,Bavota2015}. However, our analysis quantifies this influence at scale and underscores the systemic risk associated with these libraries, failures or vulnerabilities in these nodes could propagate across a vast number of dependent projects.

\begin{table}[ht]
\renewcommand{\arraystretch}{1.3}
  \centering
  \caption{To 10 artifacts in the Maven Central dependency graph with the highest PageRank scores, representing potential ecosystem hubs; includes each artifact’s identifier, functional category, and associated descriptive tags}

  \label{tab:pagerank}
  \rowcolors{2}{gray!25}{white}
  \begin{tabularx}{\linewidth}{l|Y|c|Y|Y}
  \rowcolor{white}
    \toprule
    \textbf{Rank}
      & \textbf{Node Name}
      & \textbf{PageRank Score}
      & \textbf{Category}
      & \textbf{Tags} \\
    \midrule
    1
      & org.apache.felix:org.apache. felix.scr.ds-annotations
      & 0.0043
      & Defect detection metadata
      & Quality, defect, annotations, analysis, metadata, jetbrains \\
    2
      & org.apache.felix:org.apache. felix.scr.ds-annotations:1.2.8
      & 0.0041
      & Defect detection metadata
      & Quality, defect, annotations, analysis, metadata, jetbrains \\ 
    3
      & org.apache.felix:org.apache. felix.scr.ds-annotations:1.2.4
      & 0.0041
      & Defect detection metadata
      & Quality, defect, annotations, analysis, metadata, jetbrains \\ 
    4
      & org.apache.felix:org.apache. felix.scr.ds-annotations:1.2.2
      & 0.0041
      & Defect detection metadata
      & Quality, defect, annotations, analysis, metadata, jetbrains \\ 
    5
      & org.apache.felix:org.apache. felix.scr.ds-annotations:1.2.0
      & 0.0041
      & Defect detection metadata
      & Quality, defect, annotations, analysis, metadata, jetbrains \\ 
    6
      & org.apache.felix:org.apache. felix.scr.ds-annotations:1.2.10
      & 0.0041
      & Defect detection metadata
      & Quality, defect, annotations, analysis, metadata, jetbrains \\ 
    7
      & junit:junit
      & 0.0039
      & Testing Frameworks \& Tools
      & Testing, junit, quality \\
    8
      & org.hamcrest:hamcrest-all:1.1
      & 0.0026
      & Testing Frameworks \& Tools
      & Matching, hamcrest, testing, quality \\
    9
      & org.hamcrest:hamcrest-all:1.0
      & 0.0026
      & Testing Frameworks \& Tools
      & Matching, hamcrest, testing, quality \\
    10
      & org.hamcrest:hamcrest-all:1.3
      & 0.0026
      & Testing Frameworks \& Tools
      & Matching, hamcrest, testing, quality \\
    \bottomrule
  \end{tabularx}
\end{table}

We further analyzed the dependency graph by computing betweenness centrality to identify nodes that act as critical connectors or bridges between different parts of the ecosystem. Table~\ref{tab:betweenness} presents the top 10 nodes ranked by betweenness centrality score, highlighting components that serve as key pathways for dependency flows. Our findings show that \texttt{io.micrometer:micrometer-core} occupies the highest position, indicating its role as a pivotal component for application observability and performance monitoring across the ecosystem. This is followed closely by \texttt{org.springframework:spring-core}, a foundational utility in the Spring framework that supports a wide array of applications and services. Other prominent nodes include \texttt{io.projectreactor:reactor-core}, facilitating concurrency and parallelism, and \texttt{org.mariadb.jdbc:mariadb-java-client}, a widely adopted JDBC driver enabling database connectivity. 

The presence of these nodes as top-ranked by betweenness centrality suggests they not only serve their primary purposes but also act as bridges connecting otherwise disparate modules, frameworks, or services. Interestingly, essential libraries for testing (\texttt{org.assertj:assertj-core}, \texttt{org.testng:testng}) and logging (\texttt{org.apache.logging.log4j:log4j-core}) also appear among the top bridging nodes, reinforcing their ecosystem-wide integration beyond individual projects. These results highlight the dual role of these libraries as specialized tools and as integrative backbones of the ecosystem, emphasizing their criticality from both functionality and network structure perspectives. Their high betweenness scores underline their systemic importance, where disruptions or vulnerabilities could isolate significant portions of the ecosystem or break dependency chains. Conversely, the lowest-ranked nodes by incoming dependencies shown in Table~\ref{tab:low-betweenness} are primarily API management components from the \texttt{WSO2} ecosystem, reflecting their peripheral role. These modules are specialized and domain-specific, serving fewer projects and indicating limited reuse across the broader \texttt{Maven Central} ecosystem.

\begin{table}[ht]
  \renewcommand{\arraystretch}{1.0}      
  \centering

  \caption{Ten artifacts in the Maven Central dependency network ranked by highest betweenness centrality, highlighting the key bridges that connect distinct clusters; includes each artifact’s identifier, functional category, and descriptive tags.}

  \label{tab:betweenness}
  \rowcolors{1}{gray!25}{white}  
  \begin{tabularx}{\linewidth}{%
      l     
      |Y    
      |c     
      |Y    
      |Y    
    }
    \rowcolor{white}\toprule
    \textbf{Rank}
      & \textbf{Node Name}
      & \textbf{Betweenness Score}
      & \textbf{Category}
      & \textbf{Tags} \\
    \midrule

    1 & io.micrometer:micrometer-core
      & 0.0644
      & Application Metrics
      & Observability, monitoring, management, metrics, performance \\
    2 & org.springframework:spring-core
      & 0.0604
      & Core Utilities
      & Beans, context, spring, IoC, framework \\
    3 & io.projectreactor:reactor-core
      & 0.0603
      & Concurrency Libraries
      & Reactor, concurrency, parallel, multithreading \\
    4 & com.hazelcast:hazelcast
      & 0.0465
      & In-Memory Data Grid
      & Hazelcast, clustering, caching, distributed data \\
    5 & org.mariadb.jdbc:mariadb-java-client
      & 0.0450
      & JDBC Drivers
      & Database, SQL, JDBC, driver, MariaDB, client, RDBMS, MySQL \\
    6 & org.assertj:assertj-core
      & 0.0443
      & Assertion Libraries
      & Assert, quality, testing, assertion, fluent, validation \\
    7 & org.testcontainers:mariadb
      & 0.0427
      & Database Testing Tools
      & Database, container, testing, MariaDB, MySQL \\
    8 & com.amazonaws:aws-java-sdk
      & 0.0378
      & Cloud Storage SDK
      & Persistence, AWS, S3, Amazon, SDK, client, storage \\
    9 & org.apache.logging.log4j:log4j-core
      & 0.0365
      & Logging Frameworks
      & Logging, Log4j, Apache \\
    10 & org.testng:testng
       & 0.0283
       & Testing Frameworks
       & Testing, quality assurance \\
    \bottomrule
  \end{tabularx}
\end{table}

\begin{table}[ht]
  \renewcommand{\arraystretch}{1.3}  
  \centering
  \caption{Top 10 leaf nodes in the Maven Central dependency graph, ranked by their minimal incoming dependency count. These artifacts represent end‐point packages with the fewest dependents and thus minimal downstream reuse or highly specialized functionality.}
  \label{tab:low-betweenness}
  \rowcolors{1}{gray!25}{white}
  \begin{tabularx}{\linewidth}{%
      c    
      |Y   
      |c    
      |Y   
      |Y   
    }
    \rowcolor{white}
    \toprule
    \textbf{Rank}
      & \textbf{Node Name}
      & \textbf{Incoming Edges}
      & \textbf{Category}
      & \textbf{Tags} \\
    \midrule
    
    1 
      & org.wso2.carbon.apimgt:org. wso2.carbon.apimgt.core
      & 2967
      & API Management
      & Bundle, api, osgi \\
    2 
      & org.wso2.carbon.apimgt:org. wso2.carbon.apimgt.core. feature
      & 2967
      & API Management
      & Api \\
    3 
      & org.wso2.carbon.apimgt:org. wso2.carbon.apimgt. throttling.siddhi.extension
      & 2927
      & API Management
      & Bundle, api, osgi, extension \\
    4 
      & org.wso2.carbon.apimgt:org. wso2.carbon.apimgt.keymgt. client
      & 2722
      & API Management
      & Bundle, client, api, osgi \\
    5 
      & org.wso2.carbon.apimgt:org. wso2.carbon.apimgt.api
      & 2722
      & API Management
      & Bundle, api, osgi \\
    6 
      & org.wso2.carbon.apimgt:org. wso2.carbon.apimgt.keymgt. stub
      & 2722
      & API Management
      & Bundle, api, stub, osgi \\
    7 
      & org.wso2.carbon.apimgt:org. wso2.carbon.apimgt.impl
      & 2721
      & API Management
      & Bundle, implementation, api, osgi \\
    8 
      & org.wso2.carbon.apimgt:org. wso2.carbon.apimgt.gateway
      & 2721
      & API Management
      & Bundle, gateway, api, osgi \\
    9 
      & org.wso2.carbon.apimgt:org. wso2.carbon.apimgt.keymgt
      & 2721
      & API Management
      & Bundle, api, osgi \\
    10
      & org.wso2.carbon.apimgt:org. wso2.carbon.apimgt. keymanager.feature
      & 2712
      & API Management
      & Manager, api \\
    \bottomrule
  \end{tabularx}
\end{table}

\subsubsection*{Discussion and Implications.}

Our analysis of PageRank and betweenness centrality reveals a clear structural divide within the \texttt{Maven Central} ecosystem. A small number of libraries serve as highly influential hubs, while the majority of components remain on the periphery with limited influence. Libraries such as \texttt{org.apache.felix.scr.ds-annotations, junit:junit}, and \texttt{org.hamcrest:hamcrest-all} appear at the top of the PageRank rankings. These artifacts are widely reused across many domains and often appear deep in dependency chains. Their high centrality reflects their foundational role in the ecosystem, supporting tasks like annotation processing, unit testing, and quality assurance. Similarly, libraries with high betweenness centrality such as \texttt{spring-core} and \texttt{micrometer-core} act as bridges between different parts of the graph. These components connect otherwise separate modules and facilitate interoperability between frameworks. Because they sit on many paths between other libraries, they are critical for maintaining ecosystem cohesion. However, this centrality also introduces risk. If a highly central library becomes deprecated, vulnerable, or poorly maintained, its failure can affect a large number of downstream projects. These libraries act as ``choke points,'' and disruptions can cause widespread instability, failed builds, or security exposure across the ecosystem.

On the other end of the spectrum are low-centrality nodes, many of which come from the \texttt{org.wso2. carbon.apimgt} package family. These components are domain-specific, used primarily in narrow contexts such as API gateway management. They are reused less frequently and have few connections to other parts of the graph. While their limited influence reduces the risk of widespread disruption, it also means they may lack visibility, strong community support, or long-term maintenance. These traits can make them vulnerable to neglect, even if they pose less systemic risk. These findings highlight the value of centrality-aware software engineering. Tools for dependency management, security analysis, and migration planning could prioritize highly central libraries for extra scrutiny, testing, and monitoring. At the same time, maintainers of these critical packages should be supported through practices like shared maintainership, automated release pipelines, and community funding. Software developers should also weigh the structural role of a dependency before adopting it, especially if it introduces tight coupling to a critical or fragile node in the ecosystem.

\begin{custombox}
\faLightbulbO \hspace{0.1mm} \emph{\textbf{Summary - RQ2}:
We found that projects acting as network hubs predominantly belong to core ecosystem infrastructure categories, such as core infrastructure libraries (e.g., apache felix) and testing frameworks (e.g., JUnit and hamcrest). These hubs exhibit high in-degree and PageRank, reflecting their critical role in the ecosystem’s reuse structure. In contrast, peripheral nodes characterized by low degree and often no outgoing dependencies typically represent specialized application modules or niche libraries. This central-peripheral structure highlights the layered nature of the Maven Central ecosystem, where a small set of infrastructural components supports a broad, diverse periphery of application-level projects.
}
\end{custombox}

\subsection*{RQ3: Is the graph made up of connected components with no relationship between them?}

To assess the global cohesion of the \texttt{Maven Central} dependency graph, we computed both weakly and strongly connected components (Table~\ref{tab:lcc}). The analysis shows that the graph contains a total of 24 connected components when ignoring edge directionality. The largest connected component (LCC) encompasses 1,394,930 nodes, accounting for approximately 99.81\% of all nodes in the graph. This result highlights the highly cohesive nature of the ecosystem, where nearly all projects are directly or indirectly connected through dependencies. The remaining 23 small components represent only 0.19\% of the nodes, indicating the presence of isolated or niche clusters that are not integrated into the broader ecosystem. These components may correspond to abandoned projects, experimental modules, or domain-specific tools with limited reuse.

\begin{table}[ht]
  \renewcommand{\arraystretch}{1.5}  
  \centering
  \caption{Summary of graph connectivity metrics for the Maven Central dependency network, including counts of weakly and strongly connected components, sizes of the largest components, and overall coverage percentages}
  \label{tab:lcc}
  \rowcolors{1}{gray!25}{white}
  \begin{tabularx}{\linewidth}{%
      L                          
      r
    }
    \rowcolor{white}
    \toprule
    \textbf{Metric}
      & {\textbf{Value}} \\ 
    \midrule
    Total number of connected components
      & 24 \\ 
    Size of largest connected component (LCC)
      & 1,394,930 \\ 
    Total number of nodes in the graph
      & 1,397,519 \\ 
    LCC coverage
      & 99.81\,\% \\ 
    Remaining small components (23 in total)
      & 0.19\,\% \\ 
    Number of strongly connected components (SCCs)
      & 1,120,071 \\ 
    Size of largest SCC
      & 208,270 \\ 
    \bottomrule
  \end{tabularx}
\end{table}

\nd Furthermore, the graph exhibits a large number of strongly connected components (SCCs) due to its inherent directionality. In total, 1,120,071 SCCs were identified, with the largest SCC comprising 208,270 nodes. This suggests that while the ecosystem is highly connected in a weak sense, the number of fully reciprocally reachable nodes (strong connectivity) is relatively lower and confined to specific subsets of the network. This pattern is typical of dependency graphs, where dependencies often form directed acyclic structures at the macro scale but exhibit locally dense cycles within certain project families or frameworks. Overall, these findings confirm that the \texttt{Maven Central} ecosystem is structurally cohesive, with almost all artifacts interconnected in a single giant component. However, the directionality of dependencies results in fragmented strongly connected regions, reflecting the hierarchical and modular organization of software projects.

\subsubsection*{Discussion and Implications.}
Our weak connectivity analysis shows that the Maven Central dependency graph contains a single giant component comprising 99.81\% of all artifacts. This confirms the ``giant component'' phenomenon commonly seen in complex networks, where nearly all nodes are directly or indirectly connected. This result is consistent with findings from prior work on both \texttt{Maven} and \texttt{npm} ecosystems~\cite{Decan2019,Newman:2001,barabasi2013network}, which highlight how such cohesion supports rapid reuse and broad propagation of updates and patches. In such a tightly integrated ecosystem, any change such as a performance improvement or a security fix can reach a vast number of downstream projects. This amplifies both the benefits and risks of reuse. To manage this effectively, dependency tools should incorporate algorithms that efficiently compute transitive closures and assess impact, allowing developers to anticipate the ripple effects of modifications or vulnerabilities.

Although only 23 weakly connected components lie outside the giant component, they still warrant attention. These isolated subgraphs represent just 0.19\% of the ecosystem but are often unmaintained, experimental, or domain-specific. Prior research suggests such modules may carry compatibility and security risks if reintroduced into production systems~\cite{kikas:2017}. Identifying and deprecating these isolates can help reduce technical debt and improve overall ecosystem hygiene. Repository maintainers could consider archiving or consolidating these components as part of long-term sustainability efforts. The strongly connected component (SCC) analysis paints a different picture. With over one million SCCs and the largest containing about 208,000 nodes, most of the dependency graph forms a directed acyclic structure with only local cycles. These bidirectional links often occur in framework internals or within tightly coupled module families. While small in scope, such cycles can complicate dependency resolution and increase the risk of version conflicts, commonly referred to as ``dependency hell''~\cite{wangying:2018}. Recognizing the boundary between acyclic regions and cyclic clusters offers opportunities for tool optimization. Dependency resolvers could treat strongly connected subgraphs as localized units, reducing global complexity during resolution. For example, conflict resolution efforts can be focused within these clusters without affecting the rest of the graph.

From a tooling and governance perspective, these results suggest a two-pronged approach. Global vulnerability scanning, migration planning, and usage analytics can leverage the reachability of the giant component. At the same time, localized resolution strategies can benefit from strong component decomposition. Ecosystem maintainers should prioritize audits and support for artifacts within the giant component, especially those with high centrality, while also monitoring peripheral or isolated clusters for archival or reintegration decisions.

\begin{custombox}
\faLightbulbO \hspace{0.1mm} \emph{\textbf{Summary – RQ3}:  
Our connectivity analysis of the Maven Central dependency graph identifies 24 weakly connected components, of which a single giant component contains 1,394,930 artifacts, or 99.81\% of the entire network. The remaining 23 minor components account for only 0.19\% of nodes and correspond to isolated or specialized packages. This structure confirms that Maven Central functions as a nearly universal software supply chain, enabling efficient transitive reuse and rapid dissemination of updates and vulnerability patches. At the same time, a few disconnected clusters highlight residual niche ecosystems that may require targeted maintenance or deprecation. 
}
\end{custombox}

\section{Threats to Validity}\label{sec:threats}
While our study provides valuable insights into the structural and connectivity patterns in the \texttt{Maven Central} dependency network, it is also subject to several threats to validity. To limit computational cost, we first selected 5,000 artifacts with highest degree centrality and then performed breadth‐first expansion to include all releases and libraries reachable from those seeds. While this approach captures the neighbourhood of highly connected hubs, it may underrepresent peripheral regions and small isolates. Graph sampling and seed selection strategies are known to bias degree and connectivity distributions if not carefully calibrated~\cite{Leskovec:2006:sampling}. However, given the dataset's characteristics, this bias is significantly reduced because over 99.81\% of the artifacts are part of a single giant strongly connected component. In addition, our study captures a single point in time and does not account for temporal dynamics such as artifact churn, version deprecation, or gradual integration of isolates into the giant component. Lastly, although \texttt{Maven Central} is the largest Java ecosystem repository, our findings may not generalize to other language ecosystems or to private corporate registries. Ecosystems such as \texttt{npm} or \texttt{PyPI} differ in their module publishing practices and dependency semantics \cite{Decan2019}. Thus, the prevalence of a giant component and the fragmentation into localized cycles may vary in ecosystems with different governance models or package formats. Future research should account for these limitations to enhance the generalizability of the findings.

\section{Conclusion}\label{sec:conclusion}
In this study, we conducted a comprehensive structural and connectivity analysis of the \texttt{Maven Central} dependency ecosystem using graph-theoretic approaches. By applying metrics such as degree distributions, PageRank, betweenness centrality, and connected components, we uncovered critical insights into the ecosystem's architecture. Our results show that \texttt{Maven Central} exhibits a scale-free and small-world structure with a small number of highly influential hubs supporting a vast periphery of dependent projects. This configuration promotes efficient reuse but also introduces systemic risks, as failures or vulnerabilities in central nodes could have widespread cascading effects. We also identified the presence of local clusters and a cohesive largest connected component, underscoring the tightly interwoven nature of the ecosystem. Overall, our findings contribute valuable knowledge for researchers, practitioners, and ecosystem maintainers, offering guidance for improving resilience, managing risks, and prioritizing future maintenance and security efforts. Future work will explore vulnerability propagation dynamics, variation of shortest path lengths between projects and communities, and cross-ecosystem comparisons. Furthermore, this analysis is based on a static snapshot of the repository. It would be interesting to examine the temporal evolution of connectivity to understand how new libraries join the ecosystem or become abandoned.  Moreover, further study of the strongly connected clusters could investigate the root causes of cycles and their impact on developer productivity.  Finally, correlating connectivity patterns with metrics such as download counts, issue resolution time, or security incident frequency could yield deeper insights into the trade-offs between cohesion and modularity in large software ecosystems.  
%
\bibliographystyle{splncs03}
\bibliography{references}
\end{document}